\documentclass[11pt]{article}
\usepackage{jair}
\usepackage[natbibapa]{apacite}
\let\cite\undef 

\newcommand{\citepos}[1]{\citeauthor{#1}'s (\citeyear{#1})}

\usepackage{tikz-dependency}
\usepackage[T1]{fontenc}
\usetikzlibrary{fit}
\usetikzlibrary{positioning}

\pgfdeclarelayer{bg}    
\pgfsetlayers{bg,main}  

\usepackage{graphicx}
\usepackage{subcaption}
\usepackage[many]{tcolorbox}
\usepackage{tabularx}
\usepackage{booktabs}
\usepackage{hyperref}
\usepackage{makecell}
\usepackage{float}

\usepackage{tikz}
\usepackage{standalone}
\usepackage{pgf-pie}

\usepackage{pgfplots}
\pgfplotsset{compat=newest}
\usepgfplotslibrary{groupplots}
\usetikzlibrary{pgfplots.statistics}

\usepackage{xparse}   
\NewDocumentCommand{\rot}{O{90} O{1em} m}{\makebox[#2][l]{\rotatebox{#1}{#3}}}%

\usepackage{amsthm}
\usepackage{amsmath,amsfonts}

\newcommand{\fqt}[1]{``#1''}
\newcommand{\fqtperiod}[1]{``#1.''\xspace}
\newcommand{\fqtcomma}[1]{``#1,''\xspace}

\DeclareMathAlphabet{\mathsl}{OT1}{ptm}{m}{sl}

\newcommand{\fsub}[1]{\texorpdfstring{\textsubscript{#1}}{#1}}

\definecolor{amethyst}{rgb}{0.6, 0.4, 0.8}
\definecolor{candypink}{rgb}{0.89, 0.44, 0.48}
\definecolor{mauvelous}{rgb}{0.94, 0.6, 0.67}
\definecolor{bluegray}{rgb}{0.4, 0.6, 0.8}
\definecolor{asparagus}{rgb}{0.53, 0.66, 0.42}
\definecolor{awesome}{rgb}{1.0, 0.13, 0.32}
\definecolor{mangotango}{rgb}{1.0, 0.51, 0.26}
\definecolor{persianpink}{rgb}{0.97, 0.5, 0.75}
\definecolor{raspberry}{rgb}{0.89, 0.04, 0.36}
\definecolor{raspberrypink}{rgb}{0.89, 0.31, 0.61}
\definecolor{decentgrey}{RGB}{232,232,252}
\definecolor{oceanboatblue}{rgb}{0.0, 0.47, 0.75}
\definecolor{goldenpoppy}{rgb}{0.99, 0.76, 0.0}
\definecolor{brickred}{rgb}{0.8, 0.25, 0.33}

\definecolor{decentgrey}{RGB}{232,232,252}
\definecolor{b-teal}{RGB}{0,153,136}
\definecolor{b-orange}{RGB}{238,119,51}
\definecolor{b-olive}{RGB}{153,153,51}
\definecolor{darkblue}{RGB}{0, 130, 185}
\definecolor{shallowblue}{RGB}{61, 223, 240}
\definecolor{darkyellow}{RGB}{183, 141, 36}
\definecolor{shallowyellow}{RGB}{252, 229, 80}
\definecolor{mygrey}{RGB}{204, 204, 204}

\definecolor{mongo}{RGB}{236, 142, 51}
\newcommand{\mongosquare}[1]
{{\protect\tikz\protect\draw[fill=mongo,draw=none] (0,0) rectangle ++(0.2,0.2); #1}}
\newcommand{\mongosquarel}[1]
{{\protect\tikz\protect\draw[fill=mongo!60,draw=none] (0,0) rectangle ++(0.2,0.2); #1}}
\newcommand{\mongosquarell}[1]
{{\protect\tikz\protect\draw[fill=mongo!30,draw=none] (0,0) rectangle ++(0.2,0.2); #1}}

\definecolor{royalblue}{RGB}{65,102,245}
\newcommand{\royalbluesquare}[1]{{\protect\tikz\protect\draw[fill=royalblue, draw=none] (0,0) rectangle ++(0.2,0.2); #1}}
\newcommand{\royalbluesquarel}[1]{{\protect\tikz\protect\draw[fill=royalblue!60, draw=none] (0,0) rectangle ++(0.2,0.2); #1}}
\newcommand{\royalbluesquarell}[1]{{\protect\tikz\protect\draw[fill=royalblue!30, draw=none] (0,0) rectangle ++(0.2,0.2); #1}}

\usepackage{xspace}

\newcommand{\ifsubmit}[1]{}

\newcommand{\be}{\begin{itemize}}
\newcommand{\ee}{\end{itemize}}
\newcommand{\bn}{\begin{enumerate}}
\newcommand{\en}{\end{enumerate}}
\newcommand{\bc}{\begin{center}}
\newcommand{\ec}{\end{center}}
\newcommand{\bl}{\begin{flushleft}}
\newcommand{\el}{\end{flushleft}}
\newcommand{\beq}{\begin{equation}}
\newcommand{\eeq}{\end{equation}}
\newcommand{\bq}{\begin{quote}}
\newcommand{\eq}{\end{quote}}

\newcommand{\bmp}{\begin{minipage}}
\newcommand{\emp}{\end{minipage}}

\DeclareMathAlphabet{\mathsl}{OT1}{ptm}{m}{sl}

\usepackage{xcolor}
\usepackage{csvsimple}
\usepackage{siunitx}
\usepackage{booktabs}
\usepackage[np]{numprint}
\usepackage{multirow}
\npthousandsep{,}
\newcolumntype{T}{>{\tiny}l} 
\newcolumntype{H}{>{\Huge}l} 
\usepackage[inline]{enumitem}
\setlist[description]{leftmargin=1em}
\setlist[itemize]{leftmargin=1em}
\setlist[enumerate]{leftmargin=1.5em}

\usepackage{tikz}
\usepackage{standalone}
\usepackage{tikz-qtree}
\usepackage{float}

\usepackage[normalem]{ulem}

\usepackage[many]{tcolorbox}

\usepackage{xcolor}

\ShortHeadings{Highlighting Key Parts of Moral Sparks}
{Xi,  Singh}
\firstpageno{1}

\begin{document}

\title{Moral Sparks in Social Media Narratives}

\author{%
       \name{Ruijie Xi} \email rxi@ncsu.edu \\
       \addr Social AI Lab,
       North Carolina State University\\
       \AND
       \name{Munindar P. Singh} \email mpsingh@ncsu.edu \\
       \addr Social AI Lab,
       North Carolina State University\\
       }
\maketitle
\thispagestyle{plain}
\pagestyle{plain}
\begin{abstract}

There is increasing interest in building computational models of moral reasoning by people to enable effective interaction by Artificial Intelligence (AI) agents. 
We examine interactions on social media to understand human moral judgments in real-life ethical scenarios. 
Specifically, we examine posts from a popular Reddit subreddit (i.e., a subcommunity) called \textsl{r/AmITheAsshole}, where authors and commenters share their moral judgments on who (i.e., which participant of the described scenario) is blameworthy. 
To investigate the underlying reasoning influencing moral judgments, we focus on excerpts---which we term \emph{moral sparks}---from original posts that some commenters include to indicate what motivates their judgments. 
To this end, we examine how (1) events activating social commonsense and (2) linguistic signals affect the identified moral sparks and their subsequent judgments. 

By examining over \num{24672} posts and \num{175988} comments, we find that event-related negative character traits (e.g., immature and rude) attract attention and stimulate blame, implying a dependent relationship between character traits and moral values. 
Specifically, we focus on causal graphs involving events (\textbf{c-events}) that activate social commonsense. 
We observe that c-events are perceived with varying levels of informativeness, influencing moral spark and judgment assignment in distinct ways. 
This observation is reinforced by examining linguistic features describing semantically similar c-events. 
Moreover, language influencing commenters' cognitive processes enhances the probability of an excerpt becoming a moral spark, while factual and concrete descriptions tend to inhibit this effect. 

\end{abstract}

\maketitle
\section{Introduction}
\label{sec:intro}

With the growing autonomy and ubiquity of Artificial Intelligence (AI) technologies, it is crucial to understand how ethical principles may be integrated into the design and functionality of AI agents \citep{jobin-2019-landscape}. 
Previous research underscores trust and accountability in AI agents \citep{anderson-2007-machine, rossi-2018-building, shneiderman-2020-bridging, thiebes-2021-trustworthy}. 
For example, \citep{liscio-2022-values} build ethical AI models and agents by aligning human values and reasoning with AI agents, \citep{zari-2024-performative} connect oppression and power with AI in a social context, and \citep{tallal-2022-get} analyze curricula for teaching AI ethics to identify their commonalities. 
Other research examines the perspectives of AI researchers to identify common ground and differences in AI ethics and governance \citep{zhang-2021-ethics}, establishes connections between moral obligations in AI applications \citep{louise-2023-responsible}, and advocates for the ethical design of AI \citep{vanhee-2022-ethical}. 

\begin{figure}[!htb]
\centering
    \includegraphics[clip, trim=1cm 11.5cm 14cm 2.8cm, scale=0.8]{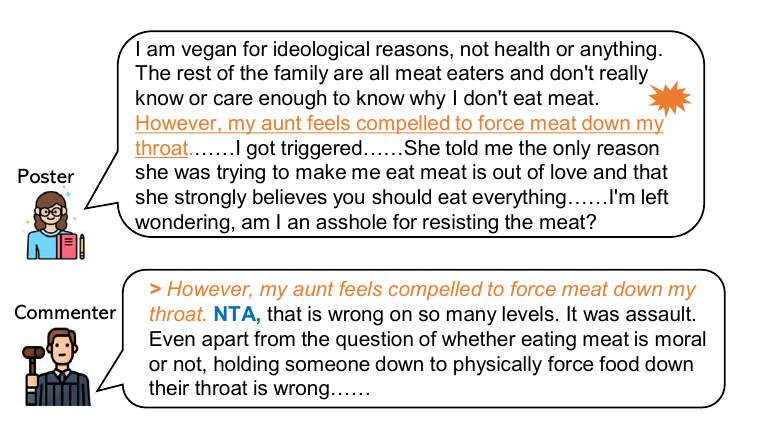}
    \caption{Example post and a comment on it that exhibits a \emph{moral spark} that in relevant to the commenter's judgment. NTA indicates the commenter stated that the author is not blameworthy.}
    \label{fig:teaser} 
\end{figure}

AI agents must model practical ethics to function effectively in real-life situations \citep{lourie-2020-scruples}. 
Researchers increasingly explore the ethical dimensions of online platforms, which provide rich real-life ethics in diverse contexts \citep{emelin-2020-moralstories, forbes-2020-social, ziems-2022-moral, shen-2022-social}. 
These platforms offer a wealth of real-life ethical scenarios and the following human moral judgments, providing researchers with valuable data to understand the complexities of ethics. 
Consider Reddit, a popular social media platform, housing numerous topic-specific subcommunities called \fqt{subreddits} and enabling diverse studies in understanding human behaviors. 
Specifically, analyzing discussions on r/AmITheAsshole (AITA)\footnote{\url{https://www.reddit.com/r/AmItheAsshole/?rdt=44617}} reveals diverse moral dilemmas, offering insights into varied ethical perspectives across cultures. 
Recent studies have investigated AITA subreddit to understand real-life moral scenarios \citep{lourie-2020-scruples, zhou-2021-assessing, nguyen-2022-mapping, nguyen-2023-measuring, xi-2023-blame, xi-2023-mundane}. 
Figure~\ref{fig:teaser} shows an example of a post and a comment it received. 
In AITA, an \emph{author} posts a story of interpersonal conflicts seeking opinions on whether they are blameworthy. 
Others---the \emph{commenters}---provide moral judgments by providing verdict codes (e.g., YTA and NTA) on who is blameworthy (i.e., the author (YTA) or others (NTA)) and provide their reasoning. 
The final verdict (judgment of who is \emph{blameworthy}) of a post is the top-voted comment's verdict. 

Moral narratives in AITA contribute to grounding descriptive ethics, enriching the comprehension of social and cultural knowledge in familiar situations. 
AITA has gained popularity in computational studies \citep{nguyen-2022-mapping, botzer-2022-analysis, xi-2023-blame, giorgi-2023-author} of social interaction that focus on comprehending moral values. 
Previous works focus on extracting social norms from AITA \citep{lourie-2020-scruples} and constructing crowdsourced datasets based on the extracted norms \citep{emelin-2020-moralstories, forbes-2020-social, ziems-2022-moral, shen-2022-social}, including the annotation of whether machine-generated dialogues are ethical \citep{ziems-2022-moral}. 

Despite the importance and popularity of learning moral values in AITA, the use of moral reasoning in AITA is underexplored. 
As shown in Figure~\ref{fig:teaser}, commenters use \fqt{$>$} to quote excerpts from the original posts, signifying focal points they respond to. 
Quoted excerpts in Reddit posts are helpful in reasoning studies such as to understand argumentation \citep{joyohan-2020-detecting}. 
However, these quoted excerpts remain unstudied in this setting. 
We refer to such a quoted excerpt as a \textbf{moral spark} since such excerpts attract moral attention and subsequently influence judgments, whether supportive or critical of the characters (author and others) in the post. 

This work seeks to deepen our understanding of moral narratives in AITA from the perspective of reasoning. 
Reasoning, as vital in narrative comprehension \citep{todorov-1981-introduction, graesser-2003-mind, piper-2021-narrative}, involves deciphering not only the literal content but also the rich nonliteral implications influenced by social, cultural, and moral conventions. 
Prior works investigate reasoning in social contexts by leveraging \emph{commonsense} knowledge, such as about behavioral intentions and consequences \citep{forbes-2020-social, emelin-2020-moralstories}. 
However, no previous work has integrated commonsense causal reasoning to study moral narratives in AITA. 
Moreover, previous works find that linguistic features affect how a commenter assigns blameworthiness in AITA \citep{zhou-2021-assessing, botzer-2022-analysis, candia-2022-demo, xi-2023-blame}. 
However, no previous work has investigated whether linguistic features affect how a commenter identifies moral sparks, which are pivotal in the moral judgments being produced. 

\begin{figure}[!htb]
    \centering
    \includegraphics[clip, trim=0cm 15cm 20cm 3cm, scale=0.8]{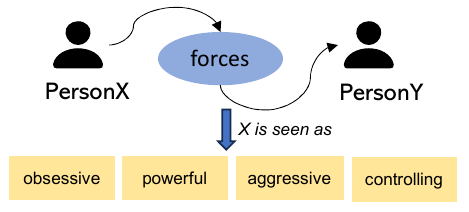}
    \caption{Example event-attribute relationships from ATOMIC \citep{hwang-2021-comet} for the moral spark in Figure~\ref{fig:teaser}.}
    \label{fig:teaser_xattr}
\end{figure}

Accordingly, we tackle the following research questions:
\begin{description}

\item[RQ\fsub{commonsense}:] \textbf{How does commonsense affect moral spark identification in AITA?} We investigate moral sparks by leveraging a knowledge graph, ATOMIC \citep{hwang-2021-comet}, which provides a causal graph involving events (\textbf{c-events}) that activate social commonsense. 
C-events in ATOMIC are structured as sentence snippets, such as of the form \fqtcomma{PersonX accuses PersonY} and are connected by different causal relation edges. 
We are interested in the \textsl{is seen as} event-attribute relation because it contains event-perceived character traits, constituting the core information conveyed to audiences \citep{genette-1983-narrative}. 
As illustrated in Figure~\ref{fig:teaser_xattr}, identifying that the event \fqt{forces} causes perceived character traits like \textsl{controlling}, which enables us to investigate which c-events and character traits contribute to a moral spark. 

\item[RQ\fsub{judgment}:] \textbf{How do moral sparks affect moral judgments in AITA?} 
We further investigate how c-events in moral sparks affect the subsequent blameworthiness assignment, thereby enriching our understanding of descriptive characters in narratives and the dynamics within narrative structures. 

\item[RQ\fsub{linguistic}:] \textbf{How do linguistic features influence moral spark identification in AITA?} We extract linguistic features of each data point based on Moral Foundation Theory \citep{haidt-2007-mft} and other cognitive models \citep{mohammad-2013-emotion, pennebaker-2015-development}. 
We cluster c-events by semantic similarity to unveil correlations between linguistic features and c-events. 
By doing so, we identify an additional layer of factors contributing to moral sparks in conjunction with similar c-events. 
\end{description}

\paragraph{Findings}

Our computational framework yields interesting findings, suggesting that there is an overall tendency when identifying moral sparks and assigning judgments in AITA. 
Specifically, descriptive character traits have varying effects in moral narratives: negative traits attract commenters' attention and amplify blameworthiness, while the traits that evoke sympathy reduce blameworthiness. 
For example, an instance featuring the c-event \fqt{PersonX gets bullied} elicits the character trait of \textsl{sad}, which is likelier to become a spark than other traits. 
This pattern can be explained by psychological research that individuals are motivated to assess character traits and not just violations of social norms, with traits offering varying levels of informativeness \citep{uhlmann-2015-person}. 
For example, character traits that evoke reader sympathy mitigate blameworthiness \citep{gray-2011-blame}. 

We additionally examine the impact of linguistic features and find that they reaffirm the importance of character traits in moral sparks. 
For instance, when the events are semantically similar, characters that are portrayed as \textsl{Powerful} (over others) and in \textsl{Negative} sentiment are likelier to attract attention. 
Also, we find that words related to a moral theory, Moral Foundation Theory \citep{haidt-2007-mft}, such as \textsl{Disgust} and \textsl{Harm} trigger moral spark identification in some domains (e.g., \textsl{Emotion-Fueled Moments}). 
Conversely, when factual information such as background (e.g., \textsl{Time}) is provided, it generally makes an instance less likely to be a moral spark. 

\paragraph{Contributions}

To the best of our knowledge, we are the first to study moral sparks in AITA via commonsense causal reasoning and linguistic features. 
Our computational method enriches comprehension of narratives. 
Our results align with psychological research, indicating that individuals evaluate moral situations by considering both social norms and descriptive character traits, gaining varying levels of informativeness from such traits \citep{uhlmann-2015-person}. 
Therefore, our findings advance research on exploring moral values from social media, urging researchers to consider the impact of descriptive character traits when investigating moral values such as constructing crowdsourced datasets of moral values based on moral narratives. 
We have released our data\footnote{\url{https://zenodo.org/record/8350526}}. 

\section{Related Work}

\subsection{Social Commonsense Causal Reasoning}

Social commonsense reasoning has gained popularity for tasks such as identifying event causality in social media \citep{kayesh-2022-deep}. 
\citet{rashkin-2018event-2mind} introduce a crowdsourced text dataset for commonsense inference between everyday events. 
\citet{emelin-2020-moralstories} construct a crowdsourced dataset including moral actions, intentions, and consequences. 
\citet{forbes-2020-social} extract Rules of Thumb from moral judgments to build commonsense rules of daily life. 
Moreover, incorporating social commonsense reasoning enhances the performance of pretrained language models \citep{chang-2020-incorporating} and facilitates the detection of cultural biases \citep{bauer-2023-social}. 

\subsection{Moral Narrative Understanding of AITA}

First-person narratives are likelier to be nonlinear and contain more salient events than other narratives \citep{sap-2022-quantifying}. 
Methods used to understand these narratives include word embeddings \citep{antoniak-2019-narrative}, topic modeling \citep{nguyen-2022-mapping}, and sentiment analysis \citep{giorgi-2023-author, xi-2023-blame}. 
Moreover, this difference raises a second problem: first-person perspectives in narratives create a distinct dynamic where authors embody both the narrator and a character \citep{bal-2009-narratology}. 
An author can subtly influence a reader's perspectives via seemingly objective statements, e.g., the choice of predicates conveying nuanced connotations \citep{rashkin-2016-connotation, sap-2017-connotation}. 
Previous works find that linguistic features affect understanding first-person narratives in social media. 
These features include the use of passive voice in blame assignment \citep{zhou-2021-assessing} and power dynamics in birth stories \citep{antoniak-2019-narrative}. 

Moral narratives in AITA exemplify descriptive ethics. 
\citet{lourie-2020-scruples} predict judgments of morality or immorality using social norms collected from AITA. 
Other works construct crowdsourced datasets \citep{emelin-2020-moralstories, forbes-2020-social, ziems-2022-moral,shen-2022-social} with a focus on social, moral, and cultural understanding based on social norms collected from \citepos{lourie-2020-scruples} dataset. 
Another line of work statistically analyzes moral narratives in AITA. 
\citet{nguyen-2022-mapping} give a taxonomy of the structure of moral discussions. 
\citet{botzer-2022-analysis} investigate how users provide moral judgments, finding that commenters prefer posts with a positive moral valence. 
\citet{candia-2022-demo} find that young and male authors are likelier to receive negative judgments, and society-relevant posts are more likely to receive negative moral judgments than romance-relevant posts. 
\citet{xi-2023-blame} separate characters as protagonists and antagonists and find evidence of bias, such as that males are likelier to receive blame. 
\citet{giorgi-2023-author} extract linguistic features of the posts and find that a positive tone reduces blame. 

\section{Data}
\label{sec:dataset}

Reddit discussions are structured as trees rooted at an \fqt{original} post; comments reply to the root or to other comments. 
We adopt \citepos{guimaraes-2021-xpost} definitions of \textsl{post} and \textsl{comment}. 
We define \textsl{instance} and \textsl{moral sparks} in light of our research objectives. 
\begin{description}
    \item[A post] is a starting point in a discussion. 
    \item[A top-level comment] is a reply to a post (not to a comment). 
    \item[An instance] is a full sentence parsed by the Stanford dependency parser \citep{chen-2014-fast}. 
    \item[A moral spark] is an instance quoted by comments. 
\end{description}
Following previous works \citep{lourie-2020-scruples, zhou-2021-assessing,nguyen-2022-mapping, xi-2023-blame}, we focus on top-level comments since other comments may not include verdicts and reasoning based on the posts. 

\subsection{Collect Data}

Existing datasets are either nonpublic \citep{zhou-2021-assessing, candia-2022-demo, botzer-2022-analysis} or inadequate for our objectives \citep{lourie-2020-scruples, nguyen-2022-mapping}. 
Consequently, we constructed our dataset using the PushShift API\footnote{\url{https://github.com/pushshift/api}} and Reddit API.\footnote{\url{https://www.reddit.com/dev/api}}
We employed rule-based filters following prior studies to ensure relevance and consistency between Reddit data and archived data from PushShift. 
We collect posts and comments based on their distinctive identifiers, as explained in the API documentation.\footnote{\url{https://reddit-api.readthedocs.io/en/latest/}} 
In this context, a comment's \texttt{link\_id} corresponds to the ID of the post it is a response to. 
We exclude deleted posts and comments (since they potentially violated AITA rules) as well as posts and comments from moderators and deleted accounts. 
We follow previous works \citep{xi-2023-blame, xi-2023-mundane} to select criteria for the minimum word count and the minimum number of comments and posts. 
Table~\ref{tab:instance_filter} shows our data-collection process. 
In the \textsl{rule-based collection stage}, we selected posts with at least $50$ words and at least $10$ top-level comments were selected to ensure quality. 
In the \textsl{comment quality filter stage}, we select top-level comments with a predefined verdict code and at least $15$ words. 
These efforts yielded \num{351067} posts and their corresponding 10.3M top-level comments from AITA, ranging from June 2013 to November 2021. 
In the \textsl{quoting comments selection stage}, we identified \fqt{$>$} in the instances using a regular expression, which resulted in \num{24672} posts. 

\subsection{Label Instances}

To answer RQ\fsub{commonsense} and RQ\fsub{linguistic}, we use the label $1$ to represent whether it is a moral spark, adn $0$ otherwise. 
To answer RQ\fsub{judgment}, we focus on how moral sparks affect moral judgments. 
To do so, we first extract characters in each post, followed by assigning labels on whether a person is blameworthy. 
Characters in the posts are identified through pronouns and names, events take the form of verb-driven structures (i.e., subject, predicate verb, object) \citep{candia-2022-demo, xi-2023-blame, giorgi-2023-author}. 
We use the Spacy\footnote{\url{https://spacy.io}} dependency parser to extract terms representing characters by identifying part-of-speech tags such as first-person pronouns (e.g., PRON and PROPN) for \textsl{I} and \textsl{me}. 

Following previous works \citep{lourie-2020-scruples, zhou-2021-assessing,candia-2022-demo, xi-2023-blame}, we apply regular expressions to extract a comment's verdict and label a moral spark based on it. 
Verdicts are defined in AITA: YTA (author is wrong), NTA (author is right), ESH (everyone is wrong), NAH (everyone is right), and INFO (more information needed), but comments may use short phrases instead of codes (e.g., \fqt{not the a-hole} for NTA). 
The regular expressions for extracting the codes and short phrases are in the \emph{Supplementary} material.\footnote{\url{https://zenodo.org/record/8350526}}
When multiple matches occur with a comment that includes transition words like \fqtperiod{but} we choose the one after such words. 
Extracted codes in negated judgments (e.g., \fqt{I do not think}) are reversed. 
To evaluate the step, we reviewed 500 randomly chosen comments, revealing 5\% false positives and 6\% false negatives. 

Given that the event-attribute relationships in ATOMIC indicate the perceived traits of subjects, we assign a blameworthy label to a moral spark based on its generated triple's subject. 
For instance, when a moral spark's subject refers to \textsl{I} and NTA is present, we assign a $0$; otherwise, we assign a $1$. 
To improve accuracy, we consider only the NTA and YTA verdicts. 
We handle passive constructions by using dependency trees when extracting the triples. 

\begin{table}[!htb]
    \centering
    \begin{tabular}{l S[table-format=6.0] S[table-format=8.0]}
    \toprule
         \bfseries Stage & {\bfseries \#Posts} & {\bfseries \#Comments}\\
    \midrule
         Rule-based collection & 351067 & 10296086 \\
         Comment quality filter & 51803 & 3675452 \\
         Quoting comments selection & 24672 & 175988 \\
    \bottomrule
    \end{tabular}
\caption{Data at each stage of our processes for selecting posts and comments. Each stage's output is the input to the next stage.}
\label{tab:event_data}
\end{table}

\begin{figure*}
    \centering
    \includegraphics[clip, trim=0.5cm 10.5cm 0cm 5cm, width=\textwidth]{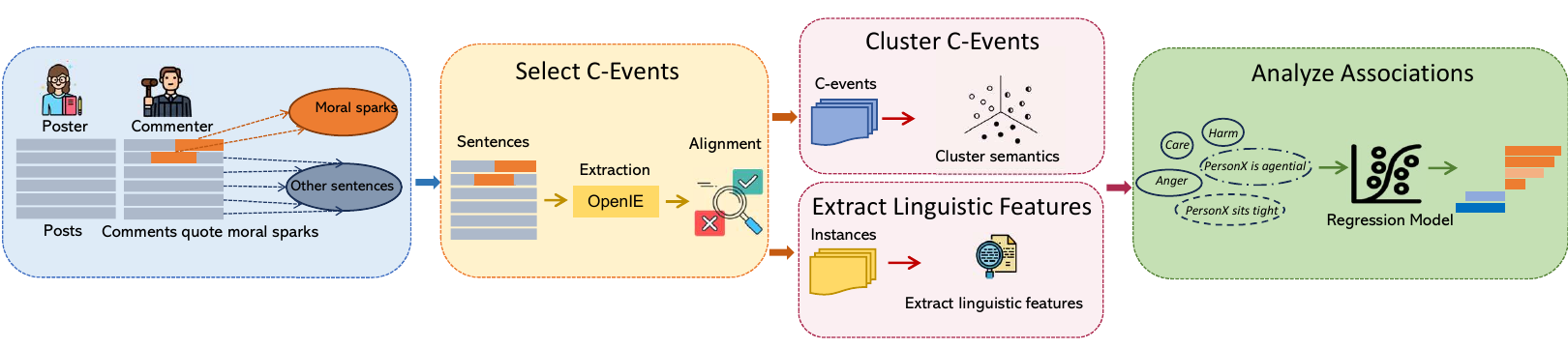}
    \caption{Our framework involves data collection, c-event selection, c-event clustering, linguistic feature extraction, and using regression models to analyze connections between moral sparks, c-events, linguistic features, and moral judgments.}
    \label{fig:framework}
\end{figure*}

\section{Methods}
\label{sec:methods}

Figure~\ref{fig:framework} shows our framework. 

\subsection{Select C-Events}
\label{sec:align_ins}

To align our data with c-events, we preprocess instances as shown in Table~\ref{tab:instance_filter}. 
We (1) parse instances from the selected posts, (2) remove unsuitable instances, and (3) align instances with c-events. 
First, we retain instances that constituency parsing recognizes as full sentences \citep{chen-2014-fast}. 
Then, we remove instances that do not contain a subject and a predicate (root) in dependency parsing. 
We improve text quality by eliminating stop words (using NLTK\footnote{https://www.nltk.org/}), emojis, and converting contractions (e.g., from \textsl{can't} to \textsl{can not}). 
And, we eliminate unicode tokens and punctuation marks for obtaining cleaner texts. 

\begin{table}[!htb]
    \centering
    \begin{tabular}{l S[table-format=6.0] r}
        \toprule
         \bfseries Stage & {\bfseries \#Instances} & {\bfseries \#Moral Sparks (\%)}\\
        \midrule
         Parse instances from \num{24672} posts &483583 &20\%\\
         Select suitable instances&318022 &21\%\\
         Align instances with c-events&228740 &20\%\\
        \bottomrule
    \end{tabular}
    \caption{Instance filtering beginning from the last line in Table~\ref{tab:event_data}.}
    \label{tab:instance_filter}
\end{table}

\begin{figure}
    \centering
    \includegraphics[clip, trim=0cm 16.8cm 21cm 3.2cm, width=0.6\columnwidth]{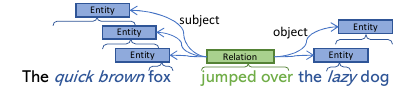}
    \caption{A COMPACTIE example. Italics are the descriptive adjectives that are part of the extracted triples.}
    \label{fig:oie_example}
\end{figure}

\paragraph{Extract events triples} 

We apply a COMPACTIE system \citep{fatahi-2022-compactie} to convert each instance into a verb-drive event triple. 
COMPACTIE yields more concise text extractions than a conventional OpenIE system. 
As shown in Figure~\ref{fig:oie_example}, extracted \textsl{(subject, predicate, object)} triples are composed of phrases or words. 

\paragraph{Align extracted triples}

ATOMIC \citep{hwang-2021-comet} provides a causal graph involving events, connecting events through nine relation edges. 
We focus on \texttt{xAttr}, which describes the perceived attributes of the subject characters. 
For instance, event \fqt{PersonX forces people} triggers PersonX \emph{is seen as} \fqtperiod{controlling, aggressive, obsessive, and powerful}

Inspired by \citet{bauer-2023-social}, we adopt two phases that balance speed and precision in this step. 
In the first phase, we employ a coarse-grained filter to efficiently collect an initial pool of ATOMIC knowledge candidates ($E_i, i=1, 2, \ldots$) for each data point ($d_i \in D$). 
We add an ATOMIC candidate event to $E_i$ if there is a verb common between $d_i$ and the candidate. 
However, we observed variations in ATOMIC data, such as \fqt{PersonX abandons the \_\_\_ altogether} and \fqtperiod{PersonX abandons \_\_\_ altogether} 
To address such variants, we generate TF-IDF vectors for each candidate and calculate the cosine similarities between every candidate in the pool. 
We identify filter candidates in the pool by setting the threshold to $0.5$. 
The selected candidates are then added to a smaller pool, denoted as $P_i$. 
The second phase prioritizes precision, assessing the semantic alignment between each ATOMIC candidate in $P_i$ and $d_i$ using BERTScore, an evaluation metric for text similarity \citep{zhang-2019-bertscore}. 
Candidates in $P_i$ are ranked based on these scores. 
Finally, we choose the three best-aligned c-events from $P_i$ for each instance. 

\subsection{Cluster C-Events}
\label{sec:cluster}

To validate the reliability of our results, we select c-events that appear at least five times in the dataset, which results in \num{2099} c-events. 
We notice that the selected c-events depict similar activities in specific scenarios, such as \fqt{PersonX visits a friend} and \fqtperiod{PersonX visits my friend}
Therefore, we cluster these c-events by semantic similarity to unveil correlations between linguistic features and the selected c-events. 
Doing so deepens the understanding of linguistic features when c-events describe similar events in different ways. 
To address the lack of external references for identifying cluster labels, we employ the Hierarchical Density-Based Spatial Clustering of Applications with Noise (HDBSCAN) algorithm \citep{mcinnes-2017-hdensity} to discover labels in an unsupervised manner. 
We adopt SentenceTransformer using Huggingface\footnote{\url{https://huggingface.co/sentence-transformers}} to encode c-events into vector representations. 
To alleviate the sparse embeddings, we perform dimension reduction using Uniform Manifold Approximation and Projection for Dimension Reduction (UMAP) \citep{mcinnes-2018-UMAP}. 
To optimize the performance of the algorithms, we fine-tune the parameters using Bayesian optimization to improve Density-Based Clustering Validation (DBCV) scores \citep{moulavi-2014-dbcv}. 
We use sklearn\footnote{\url{https://hdbscan.readthedocs.io/en/latest/}} to turn the clustered data into labels. 
Data points that are not assigned to any cluster are considered noise and labeled as $-1$: we eliminate such c-events, leaving us with a total of \num{1939} c-events in the pool. 
Our analysis is conducted on this refined c-event pool. 

\begin{table}[htb]
    \centering
    \small
    \begin{tabularx}{\linewidth}{p{1.5cm} p{4cm} X}
    \toprule
        \bfseries Category & \bfseries Feature & \bfseries Explanation \\
        \midrule
        Post & Moral Content & Occurrences of the five virtue-vice paired Moral Foundation Theory lexicon \citep{Hopp2020TheEM} \\
        Post & NRC VAD  & Occurrences of the VAD lexicon \citep{mohammad-2018-vad}\\
        Post & NRC emotion  & Occurrences of the Emotion lexicon \citep{mohammad-2013-emotion}\\
        Post & LIWC  & Occurrences of LIWC words \citep{pennebaker-2015-development}\\
        Post & Subjectivity &Occurrences of subjectivity-related words \citep{wilson-2005-recognizing}\\
        Post & Sentiment & Averaged VADER \citep{hutto-2014-vader} compound scores; nominal sentiment categories\\
        \midrule
        Cha & Connotation Frames & Scores of connotation frames-related \citep{sap-2017-connotation} words, calculated separately as writers' perspective, value, effect, and mental state\\
        Cha & Agency and Power & Agency and power scores \citep{rashkin-2016-connotation}\\
    \bottomrule
    \end{tabularx}
    \caption{Linguistic feature categories and explanations. The Post features are normalized by word counts in a sentence, where sentiment scores are not normalized; the Cha features are normalized by calculating word counts describing a character.}
    \label{tab:lin_features}
\end{table}

\subsection{Extract Linguistic Features}

Previous work showed that linguistic signals drawn from cognitive science capture the nuanced information present in moral narratives within AITA \citep{zhou-2021-assessing, xi-2023-blame, giorgi-2023-author}. 
However, no previous work has investigated them for moral sparks. 
Building upon prior research \citep{xi-2023-blame}, we create a feature score vector for each instance, and Table~\ref{tab:lin_features} presents the features of these vectors. 
Each feature vector comprises two types of features: context-level (Post) and character-level (Cha) features. 
For the Post features, we compute scores by normalizing the total number of words in each instance. 
For the Cha features, we normalize scores by the number of descriptive words associated with each character that is involved in the scenarios. 
Descriptive words are those that have grammatical relations with the character in a dependency tree (e.g., ADJ and ADV). 
We reverse the scores for a character if any negative edges (e.g., not) are found in the dependency tree. 

\paragraph{POST: Moral content} Moral Foundation Theory (MFT) \citep{haidt-2007-mft} understands how the psychological influence of social content unfolds. 
It is used to quantify moral behaviors in Twitter \citep{hoover-2020-mft} and categorize the structure of moral discussions on Reddit \citep{nguyen-2022-mapping}. 
We adopt the extended Moral Foundations Dictionary (eMFD) \citep{Hopp2020TheEM}, a crowdsourced tool for extracting moral content text, that contains \num{2041} words categorized into five domains based on MFT: care/harm, fairness/cheating, loyalty/betrayal, authority/subversion, and sanctity/degradation. 
Each word in eMFD has a composite valence score in $\lbrack-1, 1\rbrack$. 

\paragraph{POST: NRC VAD} Valence, Arousal, and Dominance (VAD) are affective dimensions to measure an author's attitudes toward the events and people referenced. 
We obtain dominance scores for \num{20000} words from the NRC VAD lexicon \citep{mohammad-2018-vad}, which contains real-valued scores from $\lbrack0, 1\rbrack$ for each category. 
This lexicon has previously been applied to moral narratives \citep{xi-2023-blame, giorgi-2023-author}. 

\paragraph{POST: NRC emotion} \citet{mohammad-2013-emotion} provide a crowdsourced list of  \num{10170} English words and their associations with eight basic emotions from \citepos{plutchik-1980-general} model (anger, fear, anticipation, trust, surprise, sadness, joy, and disgust). 
This lexicon assigns a $1$ or $0$ to each word for every emotion category, yielding an eight-dimensional vector for each instance. 
 
\paragraph{POST: LIWC} Linguistic Inquiry and Word (LIWC) organizes words into categories such as \textsl{Lifestyle} and \textsl{Psychological Processes} \citep{pennebaker-2015-development}. 
LIWC is useful in measuring moral situations in real life \citep{xi-2023-blame}. 
We use words in the expanded LIWC-22 dictionary \citep{boyd-2022-development} to vectorize word counts for each instance. 

\paragraph{POST: Subjectivity} 
We compute the subjectivity score of a post as the average score of the words in it based on \citepos{wilson-2005-recognizing} Subjectivity lexicon. 
The lexicon distinguishes subjective words as weak or strong and identifies their type as neutral, negative, or positive. 
We assign a value of $0.5$ to \fqt{weaksubj} words and $1$ to \fqt{strongsubj} words, multiplying by $0$ for neutral, $-1$ for negative, and $1$ for positive. 

\paragraph{POST: Sentiment} 
A negative tone may imply greater blame than a neutral tone. 
We determine the nominal sentiment category for each instance using VADER \citep{hutto-2014-vader}: negative for compound scores below $-0.05$, positive for scores above $0.05$, and neutral for scores in between. 

\paragraph{Cha: Connotation frames} 
We adopt \num{1000} most frequent English verbs from \citepos{rashkin-2016-connotation} dataset. 
Each verb associates four connotative dimensions in $\lbrack-1, 1\rbrack$ for a subject and a object: (1) effects, (2) values, (3) mental states, and (4) authors' perspectives. 


\paragraph{Cha: Power and agency} 
We use the power and agency dataset \citep{sap-2017-connotation}, which extends connotation frames \citep{rashkin-2016-connotation} and projects implicit power and agency levels onto individuals through their actions. 
It includes \num{1737} verbs for power and \num{2146} for agency. 
Characters with elevated agency (subjects of \textsl{attack}) assume active decision-making roles, whereas those with diminished agency (subjects of \textsl{doubts} and \textsl{needs}) are passive. 
The lexicon delineates power dynamics between subjects and objects. 
For instance, in \fqtcomma{X fears Y} Y (the object) holds power; in \fqtcomma{X pushes Y} X (the subject) wields power. 
The author's power score increases by one when they are the subject with a power label shifting from subject to object in the verb, or when they are the object and the verb's power label is shifting from object to subject. 
Agency scores are computed similar. 
Lastly, characters with power or agency over others are marked $1$, and others are marked $0$. 

\subsection{Regression Model}

We use a statistical model for measuring associations between the extracted features and moral narratives. 
For each feature, we adopt the following logistic regression model:
\begin{equation}
\log\frac{P(Y=1)}{1-P(Y=1)} = \beta_0 + \beta_X X + \sum_{i=1}^{n}\alpha_i D_i
\label{eq:1}
\end{equation}
Here, $X$ denotes the value of a characteristic (i.e., c-event or linguistic feature). 
$D_d(d = 1, \ldots , |D|)$ is a binary variable that equals $1$ if the instances belong to the $d$-th c-event domain. 
$Y$ is a binary variable that equals $1$ if an instance is a moral spark (for RQ\fsub{commonsense} and RQ\fsub{linguistic}) or a moral spark's subject is blameworthy (for RQ\fsub{judgment}). 
$\beta_X$ is the regression coefficient of characteristic $X$ and helps examine the association between the characteristic and the response; $\exp(\beta_X)$ is the odds ratio (OR) interpreted as the change of odds (i.e., ratio of probabilities) when the value of the characteristic increases by one unit. 
A positive $\beta_X$ indicates a positive association. 
To enhance reliability \citep{jafari-2019-when}, we apply a False Discovery Rate (FDR) correction \citep{benjamini-1995-controlling}. 
Since we have no a priori hypotheses, we simply examine whether $\beta_X$ has a $p$-value lower than $0.05$. 

\section{Results}
\label{sec:results}

We now examine the associations uncovered by our approach. 
Our c-event pool contains $13$ clusters, and Table~\ref{tab:clusterd_event_examples} presents the most frequent six domains, with examples for each. 

\begin{table}[!htb]
    \centering
    \small
    \begin{tabularx}{\linewidth}{X X}
    \toprule
     \bfseries Cluster&\bfseries Examples\\
     \midrule
     \multirow{2}{*}{{Actions and desires (39.3\%)}}& PersonX keeps trying\\
     {}&PersonX is able to find it\\
     \midrule
     \multirow{2}{*}{{Emotion-fueled moments (25.7\%)}}& PersonX starts screaming\\
     {}&PersonX forgets to study\\
     \midrule
     \multirow{2}{*}{{Financial situations (5.6\%)}}&PersonX pays a big fine \\ 
     {}&PersonX gets paid\\
     \midrule
     \multirow{2}{*}{Food experiences (5.2\%)}&PersonX gets fast food\\
     {}&PersonX brings a dish\\
     \midrule
     \multirow{2}{*}{Daily life (5.1\%)}& PersonX stays the night\\
     {}&PersonX comes home yesterday\\
     \midrule
     \multirow{2}{*}{Transportation (3.6\%)}&PersonX passes along the road \\ 
     {}&PersonX gets a speeding ticket\\
      \midrule
     \multirow{2}{*}{Time's passage (3.4\%)}& PersonX spends a year\\
     {}&PersonX always drank\\
     \midrule
     \multirow{2}{*}{{Romantic interactions (3.3\%)}}&PersonX supports PersonX's wife\\
     {}&PersonX becomes a couple\\
     \midrule
    \multirow{2}{*}{Life's milestones (1.8\%)}&PersonX accepts into college\\
     {}&PersonX finally started\\
     \midrule
     \multirow{2}{*}{Parenting journey (1.8\%)}&PersonX raises PersonX's children\\
     {}&PersonX gets pregnant\\
    \midrule
    \multirow{2}{*}{Hygienic (2.2\%)}&PersonX needs to use the bathroom\\
     {}&PersonX comes out in the wash\\
     \midrule
     \multirow{2}{*}{Pets love (1.4\%)}&PersonX buys a cat\\
     {}&PersonX keeps the puppy\\
     \midrule
     \multirow{2}{*}{Style and Fashion (1.3\%)}&PersonX wants a new dress\\
     {}&PersonX wears heels\\
    \bottomrule
    \end{tabularx}
    \caption{Frequent c-event clusters with two examples for each cluster.}   \label{tab:clusterd_event_examples}
\end{table}

\subsection{Human Evaluation}

We designed a survey to evaluate (1) \emph{Match}: whether each instance and its matched c-events make sense to people and (2) \emph{Name}: whether a manually created cluster name for a c-event accurately reflects the c-event's meaning. 
We randomly selected $150$ instances with their candidates from $P_i$ for Match and $10$ c-event examples from each cluster for Name. 
Three independent raters (referred to as R1, R2, and R3) provided ratings on a five-point Likert scale. 
The survey contained two questions about ratings:

\begin{description}
    \item[Match:] How closely do an event describe what happened in an instance? Please provide a rating from $1$ to $5$ with $1$ being not related and $5$ being extremely related. For example, \fqt{we got engaged last June} and \fqt{PersonX gets engaged} are extremely related. 
    \item[Name:] How accurately do a cluster name describe an event? Please provide a rating from $1$ to $5$ with $1$ being not accurate and $5$ being extremely accurate for each topic. For example, \fqt{PersonX tries to eat it} can accurately be assigned as \fqtcomma{Food Experience} but not related to \fqtperiod{Parenting Journey}
\end{description}

\begin{figure}[!htb]
\centering
\begin{minipage}[b]{0.45\columnwidth}
\centering
\begin{subfigure}[b]{\linewidth}
    \centering
    \includegraphics[width=\linewidth]{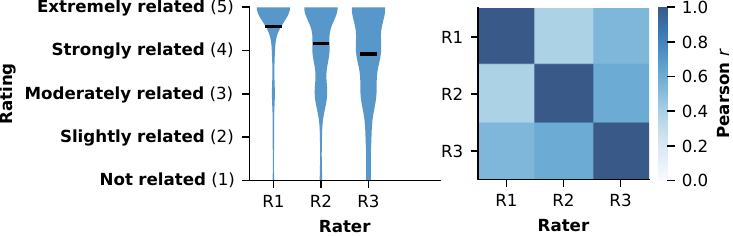}
    \caption{Survey results of Match. C-event alignments are rated on average above ``strongly related'' ($\mu=4.16$) with moderate interrater agreement ($r=0.51$).}
    \label{fig:human_eval1}
\end{subfigure}
\end{minipage}
\hfill
\begin{minipage}[b]{0.45\columnwidth}
\centering
\begin{subfigure}[b]{\linewidth}
    \centering
    \includegraphics[width=\linewidth]{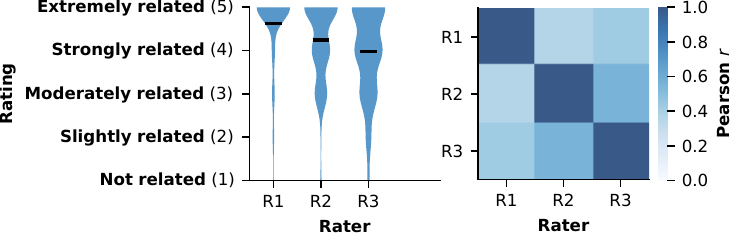}
    \caption{Survey results of Name. C-event clustering is rated on average above ``strongly related'' ($\mu=4.29$) with moderate interrater agreement ($r=0.45$).}
    \label{fig:human_eval2}
\end{subfigure}
\end{minipage}
\caption{Human evaluation of (1) match between instances and c-events and (2) quality of names given to c-event clusters.}
\label{fig:human_cluster}
\end{figure}

Figure~\ref{fig:human_eval1} and Figure~\ref{fig:human_eval2} show the results of the survey. 
The violin plots show the distributions of the ratings provided by the three raters. 
We observe that both tasks are rated above \fqt{strongly related} on average ($\mu = 4.56$ and $\mu = 4.16$). 
R3 gave the lowest  for both tasks ($\mu = 3.92$ and $\mu = 4.00$), whereas R1 gave the highest ($\mu = 4.56$ and $\mu = 4.63$). 
The heatmaps show the interrater agreements represented by Pearson correlation, demonstrating moderate agreement among the raters on average for both tasks ($r = 0.51$ and $r=0.45$). 
These results indicate that our methods of extracting and aligning social media data with commonsense knowledge are perceived as valid by humans. 

\subsection{RQ\fsub{commonsense}: Associations between C-events and Moral Sparks}
\label{sec:rqc}

\begin{figure*}[!htb]
    \centering
    \includegraphics[width=\linewidth]{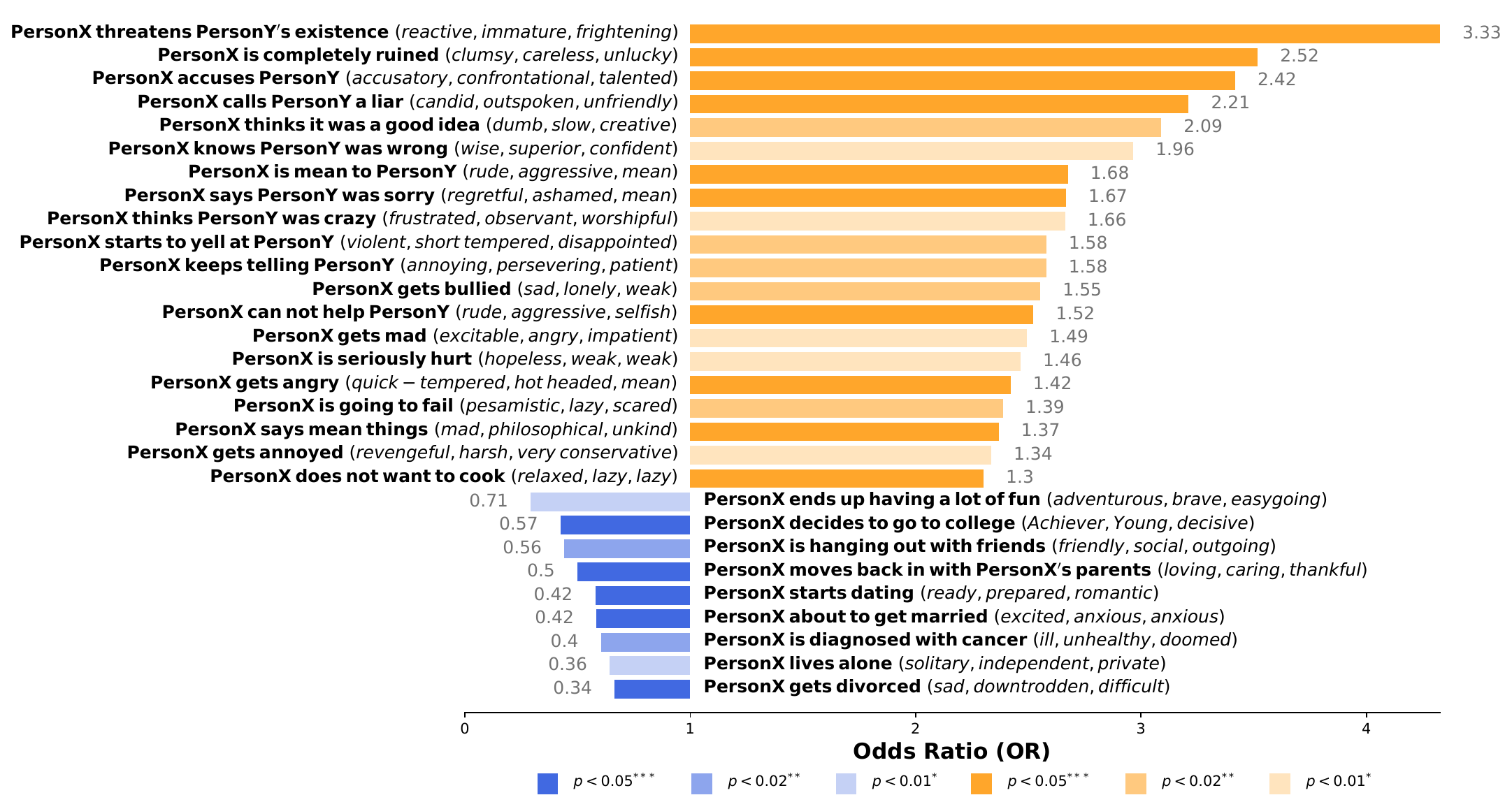}
    \caption{ The Odds Ratio (OR) values of the top $30$ frequent c-events (ordered by ORs, not frequency). An OR above one indicates that the c-event triggers a sentence being a moral spark, while an OR below one indicates the opposite. We use \mongosquare{}to indicate odds ratios greater than one and effects greater than zero (on the right), and blue rectangles \royalbluesquare{}indicate the opposite (on the left). The shade shows the $p$-values: \mongosquare{}and \royalbluesquare{} (darkest): $\le0.0001$, \mongosquarel{}and \royalbluesquarel{} (middle): $\le0.001$, \mongosquarell{}and \royalbluesquarell{}: $\le0.05$. Features are labeled using the format: \fqtperiod{Social commonsense (three random example attributes of PersonX)}}
    \label{fig:event_feats}
\end{figure*}

We estimate the effect of each event conditioned on its domain using Equation~\ref{eq:1}, and the statistical significance of the effect using the Wald test. 
We first investigate whether the c-events occurring in an instance influence if the instance becomes a moral spark. 
As depicted in Figure~\ref{fig:event_feats}, our analysis reveals a noticeable impact of c-events on moral spark identification. 
Specifically, certain c-events exhibit a positive correlation of an instance being a moral spark, increasing the likelihood from $1.42$ to $3.33$. 
We observe that c-events such as \fqt{PersonX threatens PersonsY's existence} that indicate obvious negative causal character traits (i.e., \textsl{reactive, immature, frightening} belonging to \textsl{Fear} and \textsl{Negative} emotions \citep{mohammad-2013-emotion}) trigger moral spark identifications. 
Conversely, c-events convey positive character traits such as \fqt{PersonX decides to go to college} (i.e., \textsl{achiever, young, decisive}). 
Moreover, we notice that sympathetic character traits such as \fqt{PersonX is diagnosed with cancer} (i.e., \textsl{ill, unhealthy, doomed}) could mitigate moral spark assignment. 

\begin{tcolorbox}[opacityback=0, left=2pt, right=2pt, top=2pt, bottom=5pt,arc=0pt, enhanced jigsaw, title=Takeaway from RQ\fsub{commonsense}]
C-events exhibit variability in moral spark assignment, with the activated negative character traits triggering a moral spark. 
\end{tcolorbox}

\subsection{RQ\fsub{judgment}: Associations between Moral Sparks and Judgments}

\begin{figure}[!htb]
    \centering \includegraphics[width=0.8\linewidth]{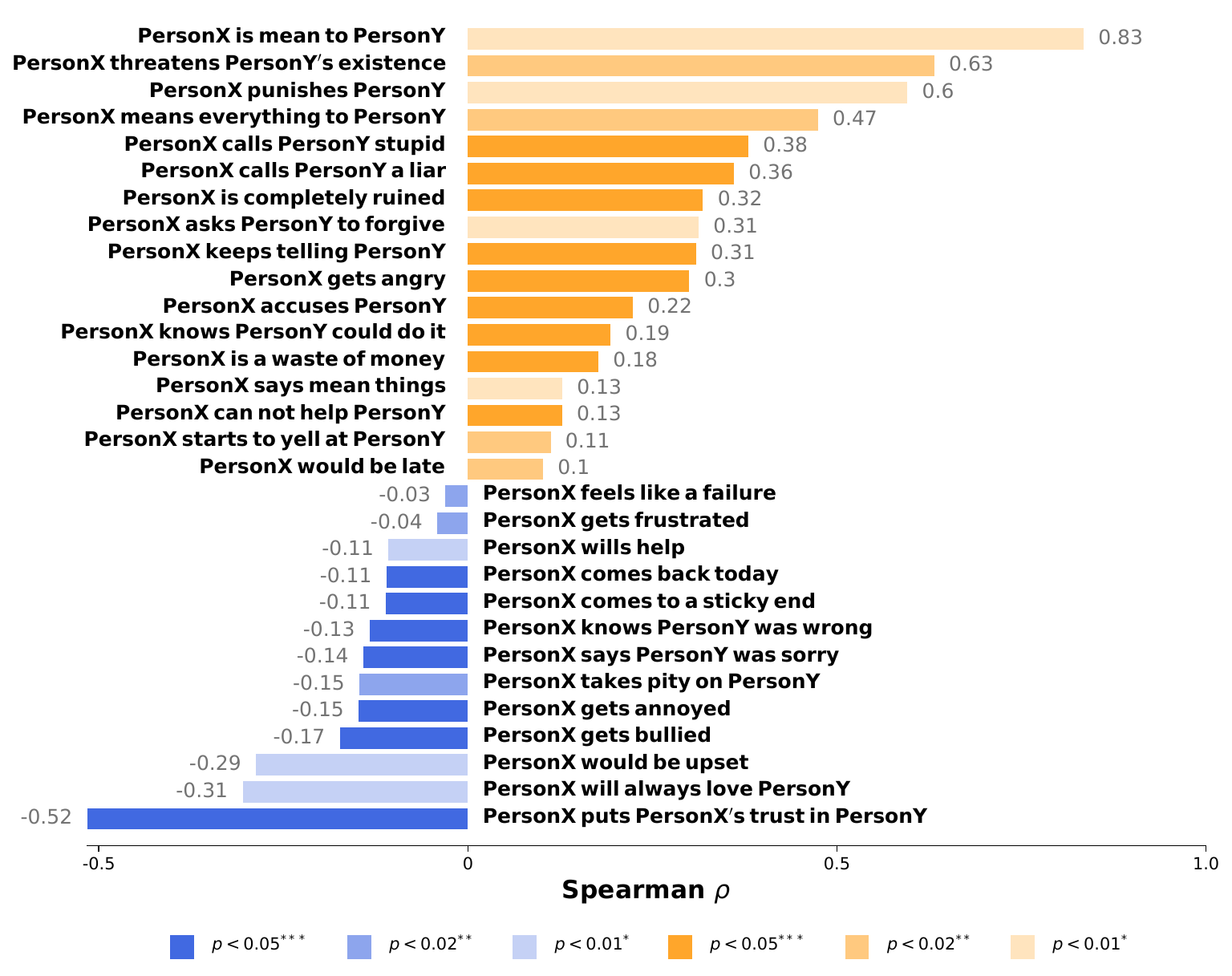}
    \caption{Correlations between moral sparks' c-events and moral judgments: 
    thirty frequent c-events with $p$-values below $0.05$ are plotted, ranked by the correlation's sign and strength.}
    \label{fig:blame_feats}
\end{figure}

\begin{figure}[!htb]
\centering
\begin{minipage}[b]{\columnwidth}
\centering
\begin{subfigure}[b]{0.45\columnwidth}
    \centering
    \includegraphics[width=\linewidth]{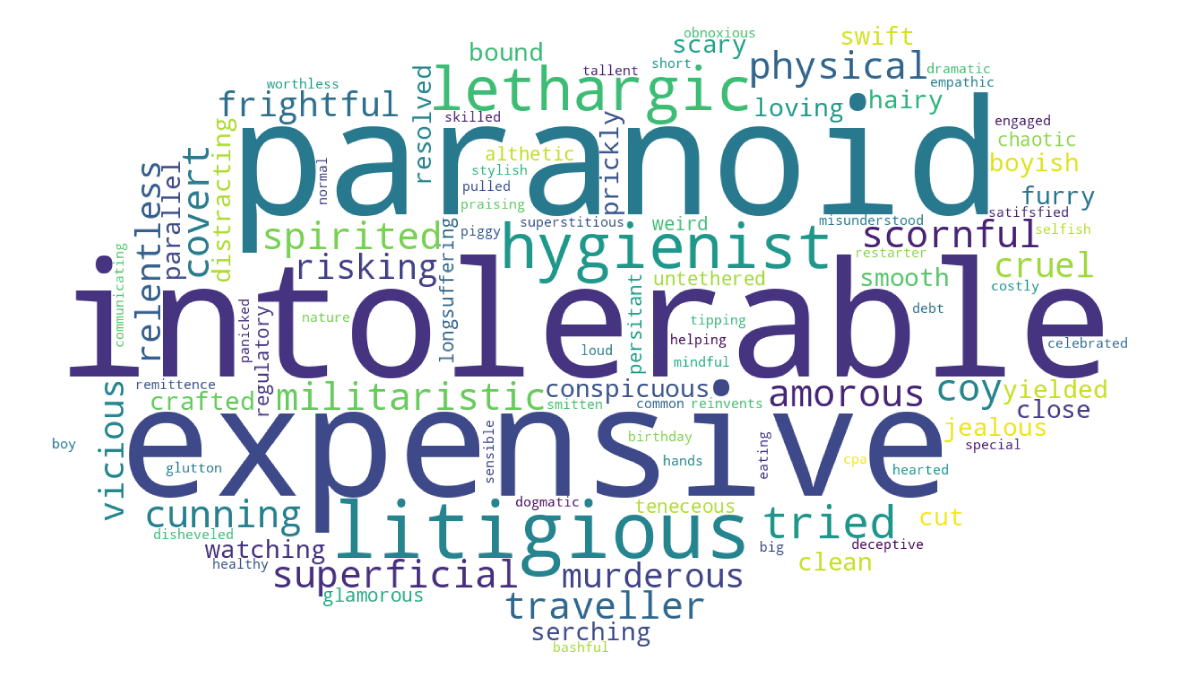}
    \caption{C-event perceived character attributes that are blameworthy.}   
\end{subfigure}
\hfill
\begin{subfigure}[b]{0.45\columnwidth}
    \centering
    \includegraphics[width=\linewidth]{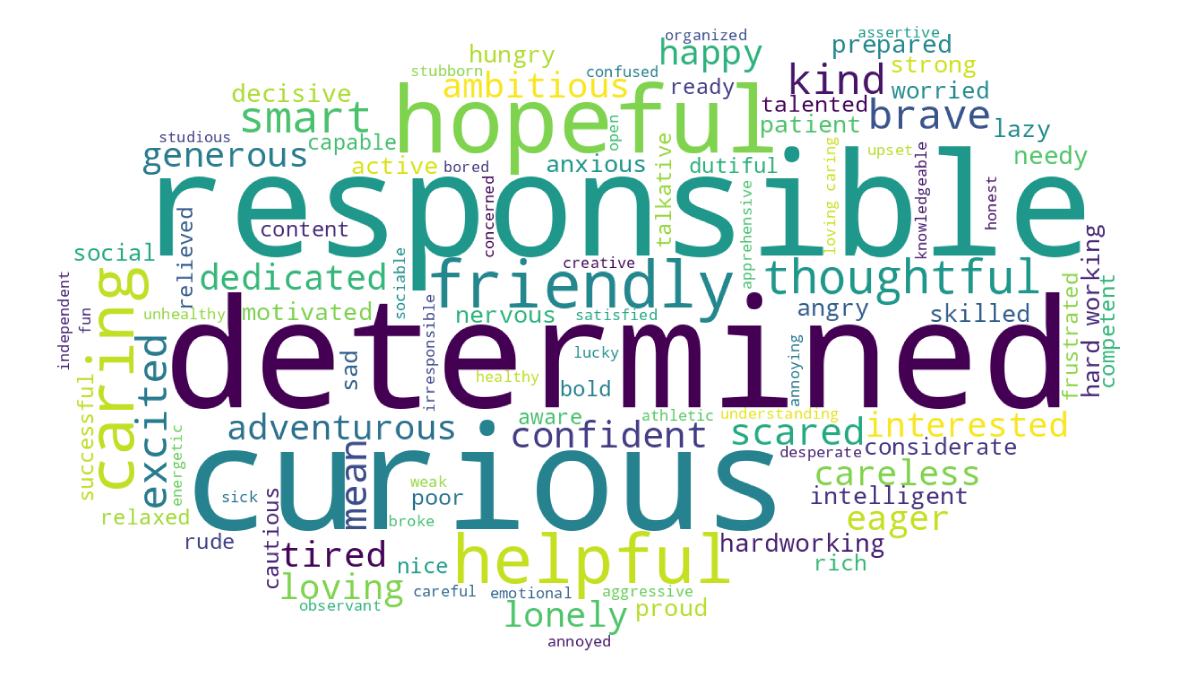}
    \caption{C-event perceived character attributes that are not blameworthy.} 
\end{subfigure}
\end{minipage}
\caption{Perceived personal attributes of c-events appearing in moral sparks.}
\label{fig:event_attrs}
\end{figure}

Figure~\ref{fig:blame_feats} depicts the Spearman correlation between c-events in moral sparks and judgments of blameworthiness. 
Higher values indicate a stronger tendency for characters involved in the c-event to be perceived as blameworthy. 
Furthermore, the attributes of a person that are signified by c-events within the c-event pool (as outlined in Section~\ref{sec:cluster}), are graphed in Figure~\ref{fig:event_attrs}. 

C-events tied to negative affects often trigger blame, illustrated by phrases like \fqt{PersonX is mean to PersonY} and \fqt{PersonX cannot help PersonY} (the attributes are shown in Figure~\ref{fig:event_feats}). 
Interestingly, we find that character traits that evoke reader sympathy trigger moral sparks but mitigate moral blame, such as feeling \textsl{sad} in the context of \fqtperiod{PersonX gets bullied} 
The observation can be explained based on social psychological findings that sympathy leads to reduced blame \citep{gray-2011-blame}. 
Moreover, we further illuminate this phenomenon by considering how linguistic features shape moral spark selection (see Section~\ref{sec:rql}). 
In addition, traits associated with negative emotions, such as the word \textsl{paranoid} in the NRC emotion \textsl{Fear} and the word \textsl{intolerable} in the NRC emotion \textsl{Anger} \citep{mohammad-2013-emotion}, amplify blame. 
These words, present in the \textsl{Psychological Process} category in LIWC \citep{boyd-2022-development}, can influence readers' comprehension of moral narratives. 
Conversely, positive attributes ascribed to a person reduce blame, such as \textsl{responsible} and \textsl{determined}. 
To conclude, our observations indicate the intricate association between moral values and character traits in moral judgments in social media narratives. 

\begin{tcolorbox}[opacityback=0,  left=2pt, right=2pt, top=2pt, bottom=5pt,arc=0pt, enhanced jigsaw,  title=Takeaway from RQ\fsub{judgment}]
C-events in moral sparks that magnify negative character traits increase blameworthiness, whereas the traits that evoke sympathy reduce blameworthiness. 
\end{tcolorbox}

\begin{figure*}[!htb]
\centering
\begin{subfigure}[b]{\textwidth}
    \centering
    \includegraphics[width=\textwidth]{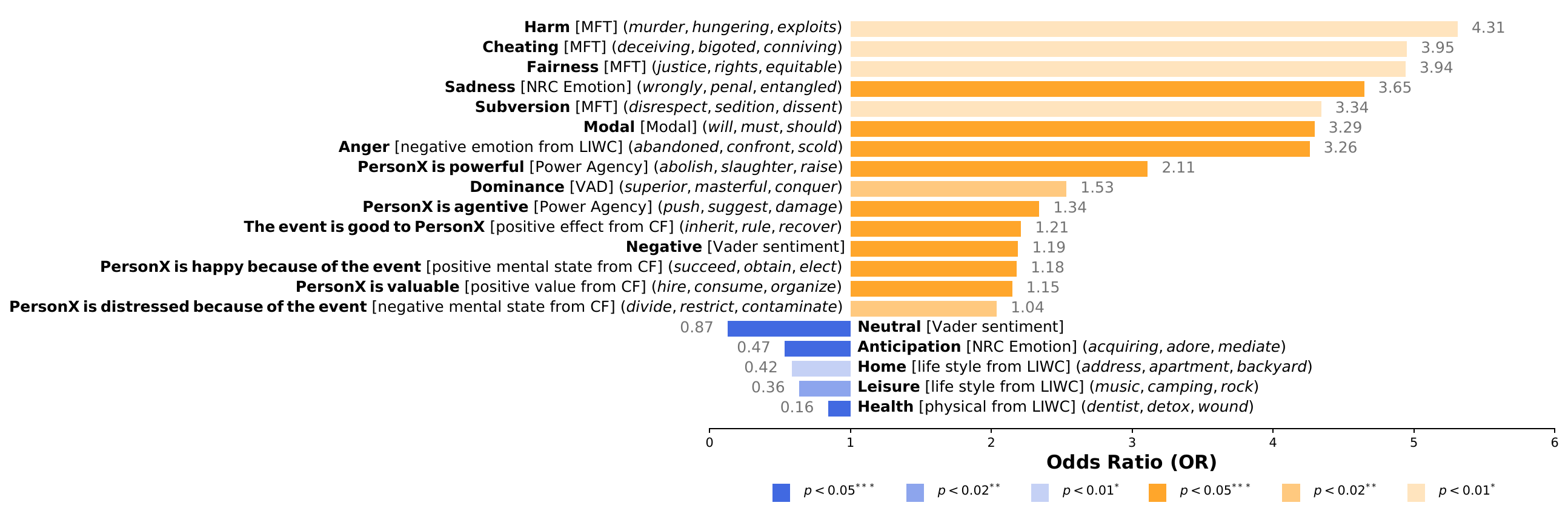}
    \caption{Actions and desires.}    
    \label{fig:coeff_action}
    \end{subfigure}
\begin{subfigure}[b]{\textwidth}
    \centering
    \includegraphics[width=\textwidth]{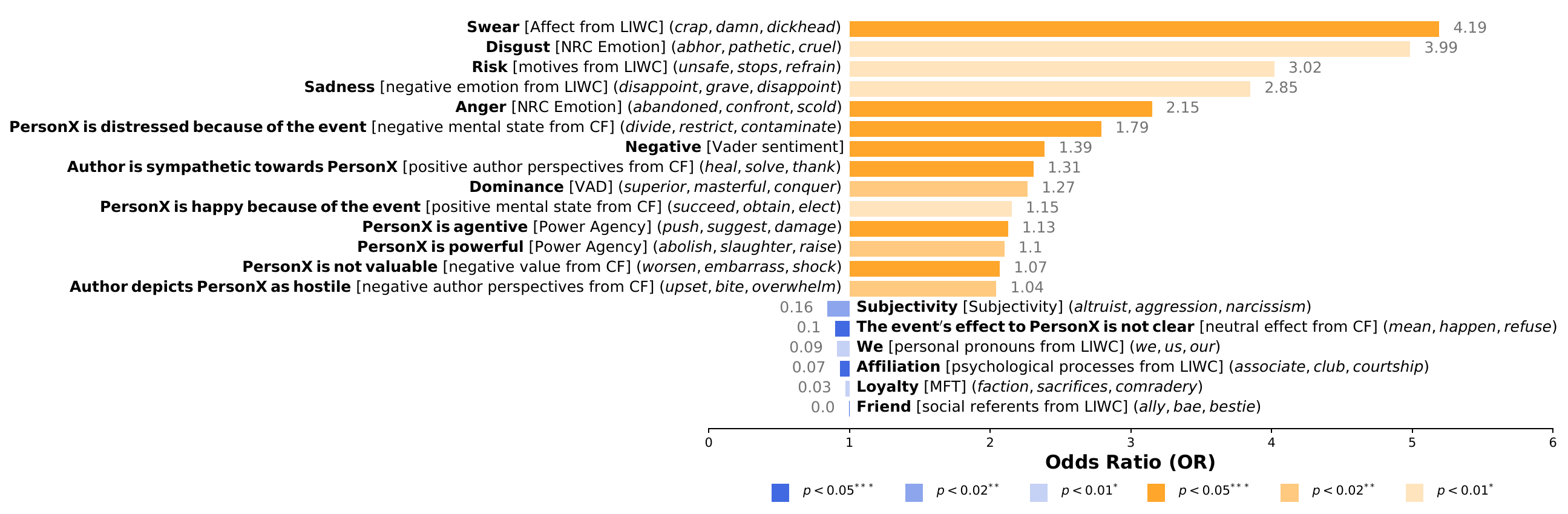}
    \caption{Emotion-fueled moments.}    
    \label{fig:coeff_emo}
\end{subfigure}
\caption{Linguistic features versus c-event domains. An OR greater than one indicates that the given feature triggers a sentence being a moral spark, while an OR below one indicates the opposite. We use orange rectangles \mongosquare{}to indicate an OR above one and effects greater than zero (on the right), and blue rectangles \royalbluesquare{}indicate the opposite (on the left). The shade shows the $p$-values: \mongosquare{}and \royalbluesquare{} (darkest): $\le0.0001$, \mongosquarel{}and \royalbluesquarel{} (middle): $\le0.001$, \mongosquarell{}and \royalbluesquarell{}: $\le0.05$. Events are labeled in the figure using the format: \fqt{Linguistic feature name [belonging lexicon]
(three random examples from the lexicon).} CF represents Connotation Frames. 
}
\label{fig:lin_or}
\end{figure*}

\subsection{RQ\fsub{linguistic}: Associations between Linguistic Features in C-Events and Moral Sparks}
\label{sec:rql}

We present the ORs and $p$-values of individual linguistic features, conditioned on c-events' domains, employing Equation~\ref{eq:1}. 
The three most frequently discussed c-event clusters are illustrated in Figure~\ref{fig:lin_or}. 
Our investigation unveils a significant trend: specific linguistic features, particularly those associated with cognitive terms, display a positive correlation with selection as a moral spark across the two domains. 
This tendency is noticeable for NRC emotion words such as the word \textsl{scold} belonging to \textsl{Anger} (OR = $3.26$, $2.15$, and $2.58$) and the \textsl{Negative} sentiment (OR = $1.19$, $1.39$, and $1.08$). 
Conversely, the \textsl{Neutral} sentiment (e.g., OR = $0.87$ in Actions and Desires) reduces the possibility of an instance being a moral spark. 
Furthermore, NRC emotion features exert distinct impacts on selection as a moral spark across different contexts, like \textsl{Sadness} on Actions and Desires (OR = $3.65$) and \textsl{Disgust} on Emotion-Fueled Moments (OR = $3.99$). 
Notably, \textsl{Swear} words in instances greatly increase moral spark identification (e.g., OR = $4.19$), potentially reflecting the authors' negative emotional states \citep{lin-2012-tweet}. 
This observation resonates with findings in Section~\ref{sec:rqc}, where heightened moral spark identification correlates with reduced positive sentiment. 

Words rooted in moral theories yield diverse effects across the domains. 
For instance, within Actions and Desires, five MFT categories exhibit positive correlations with moral spark identification, yielding ORs ranging from $3.44$ to $4.31$. 
Interestingly, \textsl{Loyalty} words show a negative correlation with moral spark identifications in Emotion-Fueled Moments (OR = $0.03$). 
Moreover, dominant tones attract moral attention. 
This is evident in the Cha features characterizing individuals as \textsl{Agentive} (e.g., OR = $1.34$ in Actions and Desires) and \textsl{Powerful} (e.g., OR = $1.1$ in Emotion-Fueled Moments), heightening the likelihood of a c-event being selected as a spark. 
The impacts of \textsl{Modal} (e.g., OR = $3.29$ in Emotion-Fueled Moments) and \textsl{Dominance} (e.g., OR = $1.29$ in Actions and Desires) words further support this finding. 
Nevertheless, the presence of \textsl{Subjectivity} diminishes moral sparks, possibly due to post-level variations. 

On a contrasting note, c-events conveying concrete details, such as \textsl{Time} and \textsl{Affiliation} (OR = $0.07$ in Emotion-Fueled Moments), exhibit reduced possibilities of moral spark identifications. 
Our observations suggest that commenters are more inclined to focus on words that bear cognitive significance than concrete descriptors when contextual details (e.g., locations of incidents) are introduced. 
These findings underscore that linguistic signals provide an additional understanding beyond social commonsense, to unravel the intricacies of moral narratives. 

\begin{tcolorbox}[opacityback=0, left=2pt, right=2pt, top=2pt, bottom=5pt,arc=0pt, enhanced jigsaw,  title=Takeaway from RQ\fsub{linguistic}]
Linguistic features modulate moral sparks: moral and cognitive words amplify the identification of a moral spark but factual and concrete descriptions inhibit it. 
\end{tcolorbox}

\section{Discussion and Conclusion}
\label{sec:conclusion}

\subsection{Main Findings}

Our study stands apart from previous works 
in exploring moral values in AITA narratives by shifting the focus from assessing whether a commenter assigns blameworthiness to examining the underlying reasoning. 
Specifically, we focus on investigating moral sparks, which are the quoted excerpts in moral narratives that attract a reader's moral attention and influence the resulting moral judgment. 
By leveraging commonsense causal reasoning, our results suggest an overall tendency, addressing both moral attention (RQ\fsub{commonsense}) and judgments (RQ\fsub{judgment}): descriptive negative character traits elicit moral attention and amplify blameworthiness, whereas traits that evoke sympathy reduce blameworthiness. 

Our results align with psychological research that individuals are motivated to assess character traits and not just violation of social norms, where the traits have varying informativeness \citep{uhlmann-2015-person} such as that character traits that evoke sympathy help mitigate blame \citep{gray-2011-blame}. 

And, we find that linguistic signals reaffirm that character traits, such as the connotative information assigned to characters, affect moral spark identification (RQ\fsub{linguistic}). 
Individuals described as \textsl{Agentive} and \textsl{Powerful}  (over others) capture a reader's focus. 
Conversely, factual and concrete (e.g., \textsl{Time}) information tends to diminish readers' engagement with moral considerations in specific scenarios such as in Actions and Desires and Emotion-Fueled Moments. 
Similarly, moral sparks that unveil negative event-perceived character attributes like \textsl{Fear} and \textsl{Anger} amplify blameworthiness. 
Furthermore, words linked to MFT \citep{haidt-2007-mft} influence moral sparks and blame in different ways. 
In addition, emotional words \citep{mohammad-2013-emotion} influence the cognitive processes governing moral focus. 
The results are aligned with research in social psychology, which highlights a strong association between words conveying cognitive effects and moral values \citep{lin-2012-tweet, sap-2017-connotation, xi-2023-blame}. 

\subsection{Broader Implications}

Our research responds to a recent call for narrative understanding grounded in causal reasoning \citep{piper-2021-narrative}. 
By employing social commonsense, we underscore the importance of unraveling the intricate web of relationships between characters and events in storytelling. 
Moreover, employing computational methods, our approach enhances the understanding of narrative structures and offers unique insights into intricate connections between social context and language. 
This aligns with previous narrative studies emphasizing the significance of reasoning in comprehending narratives \citep{todorov-1981-introduction, graesser-2003-mind, piper-2021-narrative}. 
Our methods can be applied to any social media platform featuring narrative-style posts, facilitating narrative comprehension. 
This underscores the relevance of our work in illuminating the nuanced interplay between characters, events, and their sociolinguistic context in real-world applications. 

Many studies concentrate on extracting social norms from real-life situations to construct crowdsourced datasets for downstream tasks such as improving machine ethics \citep{lourie-2020-scruples, emelin-2020-moralstories, forbes-2020-social, ziems-2022-moral, shen-2022-social}. 
However, it is crucial to note that, after when making moral evaluations, individuals do not ask themselves \fqt{is this act right or wrong?} but rather \fqt{is this person good or bad?} \citep{uhlmann-2015-person}. 
Consequently, our findings offer valuable insights for advancing research on the evaluation of moral values derived from social media data. 

\subsection{Limitations and Future Work}

Future work could enable analysis of first-person narratives in other similarly-themed social media forums such as the r/relationship\_advice subreddit. 
Although we validated the processes of aligning c-events and clustering c-events, the methods that automatically generate \textsl{(subject, predicate, object)} triples and moral judgments' labels need further validation. 
We restricted our analysis to the best-aligned c-events with each instance, which may have excluded useful information. 
Moreover, we focused solely on moral sparks but side-stepped other crucial societal factors that may influence moral judgments. 
Though we used a variety of linguistic features, we acknowledge the possibility of other features being relevant as well. 
One unexplored data source on Reddit is the comments. 
Future work could compare the comments to understand readers' comprehension of a narrative. 

\subsection{Ethical Considerations}

Reddit is a prominent social media platform. 
We scraped data from a subreddit using Reddit's publicly available official API and PushShift API, a widely used platform that ingests Reddit's official API data and collates the data into public data dumps. 
None of the commenters' information was saved during our analysis. 
The human evaluation mentioned was performed by colleagues. 

\subsection{Acknowledgment}
Thanks to the NSF (IIS-2116751) for partial support. 

\bibliography{Ruijie}
\bibliographystyle{apacite}

\end{document}